\documentclass[twocolumn,showpacs,amssymb,aps, prd]{revtex4-1}

\usepackage[usenames]{color}

\usepackage[utf8x]{inputenc}

\usepackage{graphicx}
\usepackage{epstopdf}
\usepackage{comment}

\usepackage{hyperref}
\usepackage[english]{babel}

\begin{document}

\title{Scaling up the extrinsic curvature in gravitational initial data }
\author{Shan Bai}\email{sbaiamss@yahoo.com}
\affiliation{Physics Department, University College Cork, Cork, Ireland;\\
Academy of Mathematics and Systems Science, Chinese Academy of Sciences, Beijing, China}

\author{Niall {\'O} Murchadha}\email{niall@ucc.ie}
\affiliation{Physics Department, University College Cork, Cork,
Ireland}

\date{\today}

\begin{abstract}
Vacuum solutions to the Einstein equations can be viewed as the interplay between the geometry and the gravitational wave energy content. The constraints on initial data reflect this interaction. We assume we are looking at cosmological solutions to the Einstein equations so we assume that the 3-space is compact, without boundary. In this article we investigate, using both analytic and numerical techniques, what happens when the extrinsic curvature is increased while the background geometry is held fixed.   This is equivalent to trying to magnify the  local gravitational wave kinetic energy on an unchanged background. We find that the physical intrinsic curvature does not blow up. Rather the local volume of space expands to accommodate this attempt to increase the kinetic energy.

\end{abstract}



\maketitle

\section{Introduction}
Initial data for the Einstein equations is usually constructed by the conformal method. One is given
`free' data and rescales it to get the physical initial data. This is necessary because the Einstein initial data has constraints. A comprehensive discussion of the constraints can be found in \cite{CB}, especially in Chapter VII. Interesting physics tends to occur at the boundaries of the space of free data: one gets at the very least some insight into the limitations of the conformal method.

The free data consists of a `base' metric, a Riemannian 3-metric, a divergence-free, trace-free symmetric tensor ( a TT tensor), and a scalar which is the trace of the extrinsic curvature. If the domain is a compact manifold without boundary we often choose the scalar as a constant. We then use a conformal transformation to solve the constraints. If neither the TT tensor nor the constant vanish, we can always find the appropriate conformal factor \cite{OMY}.  Parts of the boundary of the free data are easily accesible. We can scale any one of the three parts by multiplying it by a constant and letting the constant become either large or small. What effect has this on the physics? For example, does the solution just cease to exist, does the volume of the spacetime blow up (or shrink to zero), do apparent horizons (which may be interpreted as cosmological horizons) appear?

Maxwell's equations also have constrained initial data. By choosing the Maxwell free data as a pair of 3-vectors, $(\vec{A}, \vec{F})$, a parallel can be seen between the two fields. The magnetic and electric fields can be generated via $\vec{B} = \nabla\times\vec{A}$ and $\vec{E} = \vec{F} - \vec{\nabla} V$, where $V$ is a scalar function chosen to satisfy $\nabla^2V = \nabla_iF^i$. It is clear that if we multiply either $\vec{A}$ or $\vec{F}$ by any constant, $\vec{B}$ or $\vec{E}$ will be multiplied by the same amount. As a result, this kind of rescaling can be used to increase the electromagnetic energy density $(E^2 + B^2)$ without limit.

Can we perform such a rescaling in General Relativity? Can we increase the gravitational wave energy at will by multiplying  any part of the free data by a large constant?  Because the gravitational wave energy interacts in a very nonlinear way with the geometry, it is not clear what happens with the Einstein equations.

In this article, we discuss one such rescaling of the gravitational free data. The initial data consists of a 3-metric, $g_{ij}$, and a symmetric tensor, $K^{ij}$ which is the extrinsic curvature of the slice. This is essentially the velocity of the 3-metric. $K^{ij}$ is generated from the TT tensor and the constant. The constant is the Hubble constant, it represents a uniform expansion or contraction of the space, while the TT tensor (which has only 2 degrees of freedom per space point) can be interpreted as  the gravitational wave velocity. We multiply the term that represents the gravitational wave velocity by a large number and track the solution. It turns out that the local energy density remains constant, while the volume of the universe expands.

This article is a combination of analytic and numerical works. Although we study a highly nonlinear elliptic equation, nevertheless we obtain some interesting analytical results. In addition, we can accurately solve the equation numerically. This numerical work not only confirms the analytic results, where available,  but also indicates the behaviour of solutions in situations where we cannot prove anything.

In Section II we confine ourselves to describing the conformal method of solving the constraints. In Section III we show that if the extrinsic curvature vanishes nowhere on a compact manifold and we increase it, the conformal factor uniformly blows up. However, when the extrinsic curvature vanishes somewhere, the analysis in Section III is no longer valid. To investigate this special case, we revert to a spherically symmetric toy model, deriving some analytic results in Section IV and showing the numerical work in Section V. We distinguish between data sets where the extrinsic curvature vanishes in finite regions and when it only vanishes at an isolated point. Here we supply strong evidence that we do not get blow-up in regions of vanishing extrinsic curvature, and that we get slow blow-up when the extrinsic curvature vanishes at an isolated point. We conclude with a summary and an outline of future work.

  \section{Solving the Einstein Constraints}
  Initial data for the Einstein equations consists of two parts: the first part is a manifold equipped with a Riemannian 3-metric $g_{ij}$, and the second is a symmetric tensor $K^{ij}$ on the same manifold. $K^{ij}$ is the extrinsic curvature of the 3-slice, i.e., the time derivative of the 3-geometry. The metric and extrinsic curvature cannot be chosen arbitrarily. They must satisfy two constraints: the first is the Hamiltonian constraint,
  \begin{equation}
  R-K_{ij}K^{ij} + K^2= 0, \label{H}
  \end{equation}
  where $R$ is the 3-scalar curvature of $g_{ij}$ and $K$ is the trace of $K^{ij}$, i.e., $K = g_{ij}K^{ij}$. The second, the so-called momentum constraint, is
  \begin{equation}
  \nabla_iK^{ij} - \nabla^jK = 0.\label{M}
  \end{equation}
  The terminology and notation comes from \cite{ADM}.

  The standard way of generating solutions to these equations is by means of a conformal transformation. On any given manifold, it is easy to construct TT tensors \cite{JY}. These are tensors that are both tracefree and divergencefree, $g_{ij}K^{ij}_{TT} = 0; \nabla_iK^{ij}_{TT} = 0$. Such $TT$ tensors are conformally covariant. If one multiplies the given base metric $g_{ij}$ by an arbitrary positive function $\phi$ to construct a new metric , $\bar{g}_{ij} = \phi^4g_{ij}$, then $\bar{K}^{ij}_{TT} = \phi^{-10}K^{ij}_{TT}$ is TT with respect to $\bar{g}$. \cite{JY}

  Any $K^{ij}$ which is the sum of a $TT$ part and a constant trace, i.e., $K^{ij} = K^{ij}_{TT} + \frac{1}{3}Kg^{ij}$, where $K$ is a constant, satisfies the momentum constraint Eq.(\ref{M}). If we make a conformal tranformation of the metric, $\bar{g}_{ij} = \phi^4g_{ij}$, we find that $\bar{K}^{ij} = \phi^{-10}K^{ij}_{TT} + \frac{1}{3}K\bar{g}^{ij}$ again satisfies the momentum constraint with respect to the new metric. Note that $K$ is not transformed; rather, it remains a given constant. In the conformal method, one uses the fact that we can freely choose the conformal factor to solve the Hamiltonian constraint.

  If we conformally transform the metric, $\bar{g}_{ij} = \phi^4g_{ij}$, we find that the scalar curvature transforms as
  \begin{equation}
  \bar{R} = \phi^{-4}R - 8 \phi^{-5}\nabla^2\phi. \label{barR}
  \end{equation}
  We want the final metric and the final extrinsic curvature to satisfy the Hamiltonian constraint Eq.(\ref{H}), $\bar{R} - \bar{K}^{ij}\bar{K}_{ij} + K^2 = \bar{R} - \bar{K}^{ij}_{TT}\bar{K}^{TT}_{ij} + \frac{2}{3}K^2 = 0$. This reduces to the famous Lichnerowicz-York equation \cite{L, JY}
  \begin{eqnarray}
\nabla^2\phi -\frac{R}{8}\phi + \frac{1}{8}  K^{TT}_{ij}K^{ij}_{TT}  \phi^{-7} - \frac{K^2}{12}\phi^5&=& 0\\
\nabla^2\phi -\frac{R}{8}\phi + \frac{1}{8}  A^2 \phi^{-7} - \frac{K^2}{12}\phi^5&=&0\label{LY}
\end{eqnarray}
where $A^2= K^{TT}_{ij}K^{ij}_{TT}$.

This equation is very well behaved \cite{OMY}. In this article we focus on the situation where the topology of the 3-manifold is compact and without boundary. It can be shown that Eq.(\ref{LY}) has a unique positive solution if $K\ne 0$ and if $K^{ij}_{TT}$ is not identically zero. It need not be non-zero everywhere; it is enough that it not vanish somewhere \cite{OMY}. Let us remind the reader that $K$ is a constant, while $K^{ij}_{TT}$ is a function. In the special cases, where either $K = 0$ or $K^{ij}_{TT} \equiv 0$, we have an extra condition related to the sign of the scalar curvature. However, in the general case no such restriction applies. This existence result does not depend either on the metric, other than it be uniformly elliptic, or on the topology of the 3-manifold.

To recapitulate: we start with a triplet, i.e., the free data,  $(g_{ij}, K^{ij}_{TT}, K)$, and construct a new set $(\bar{g}_{ij}, \bar{K}^{ij}_{TT}, K) = (\phi^4g_{ij}, \phi^{-10}K^{ij}_{TT}, K)$ that satisfies the constraints.

We would like to investigate the boundary of the set of free data. There exist parts of the boundary that are easy to reach. We can change the original triplet in a simple way by multiplying any one of them by a constant without touching the other two, because, for example, multiplying a TT tensor by a constant does not change its TT'ness with respect to a fixed metric. Of course, such a transformation will change the solution, $\phi$, of Eq.(\ref{LY}), and thus the data satisfying the constraints will be different.

In this article we will consider such a rescaling. We pick one triplet $(g_{ij}, K^{ij}_{TT}, K)$, and use it to construct a family of free data of the form $(g_{ij}, \alpha^{12}K^{ij}_{TT}, K)$, where $\alpha$ is a running parameter. In particular, we wish to focus attention on the behaviour of $\phi$ as $\alpha$ becomes large in order to see what happens to the physical initial data that emerges. This is a particularly interesting part of the boundary.

One can regard $K^{ij}_{TT}$ as the velocity of the gravitational waves coded into the initial data.
This rescaling seems to push up the gravitational wave energy content of the free data.
Therefore, we are trying to increase the gravitational wave energy while  controlling the rest of the geometry as much as we can.

 In the special case where $K^{ij}_{TT}$ vanishes nowhere, we can show that the conformal factor uniformly blows up. We demonstrate this in the next section.

\section{a Harnack type inequality for the conformal factor}
We want to look at Eq.(\ref{LY})
$$
\nabla^2\phi -\frac{R}{8}\phi + \frac{1}{8}  A^2 \phi^{-7} - \frac{K^2}{12}\phi^5 = 0,
$$
where $A^2= K^{TT}_{ij}K^{ij}_{TT}$. We are interested in the situation where we scale $K^{TT}_{ij}K^{ij}_{TT}$ by a constant $\alpha^{12}$. Therefore we look at
\begin{equation}
\nabla^2\phi -\frac{R}{8}\phi + \alpha^{12}\frac{1}{8}  A^2 \phi^{-7} - \frac{K^2}{12}\phi^5 = 0,\label{LY1}
\end{equation}
and we wish to show that $\phi$ scales linearly with $\alpha$ as $\alpha$ becomes large. We write $\tilde\phi = \phi/\alpha$ and then Eq.(\ref{LY1}) becomes
\begin{equation}
\nabla^2\tilde\phi -\frac{R}{8}\tilde\phi + \alpha^{4}(\frac{1}{8}  A^2 \tilde\phi^{-7} - \frac{K^2}{12}\tilde\phi^5) = 0.\label{LY2}
\end{equation}
We wish to solve the family of equations on a compact manifold without boundary. It turns out that the sign of the scalar curvature plays a minor role in the behaviour of the solutions. We can always set the scalar curvature to a constant value because of the Yamabe theorem \cite{Y}, which tells us that any Riemannian metric on a compact manifold can be conformally transformed to a metric of constant scalar curvature (this really only makes sense on a compact manifold). The key quantity is the Yamabe number
\begin{equation}
Y = \inf \frac{\int[(\nabla \theta)^2 + \frac{1}{8}R\theta^2]dv}{[\int \theta^6 dv]^{1/3}},
\end{equation}
where the infimum is taken over all smooth functions, $\theta$. The sign of the Yamabe number fixes the sign of the constant scalar curvature.
Since Eq.(\ref{LY2}) is conformally covariant, and since conformal transformations form a group under composition, we can set $R$ to a constant value without losing any generality. However, we do need to handle the three separate cases, $Y > 0, R > 0, \ \ Y < 0, R < 0,$ and $Y = 0, R = 0$ independently. In each case we will set the value of $K^2 = 9$. This choice does not change in any fundamental way the behaviour of the solution.

\subsection{ $Y > 0,\ \   R > 0$}
We assume that we are in the positive Yamabe class and set the scalar curvature $R = +24$; the specific number can be chosen freely. Now Eq.(\ref{LY2}) reduces to
\begin{equation}
\nabla^2\tilde\phi - 3\tilde\phi + \alpha^{4}(\frac{1}{8}  A^2 \tilde\phi^{-7} - \frac{3}{4}\tilde\phi^5) = 0.\label{LY3}
\end{equation}
Eq.(\ref{LY3}), because it is just a rescaled version of the original Lichnerowicz-York equation Eq.(\ref{LY}), which is extremely well behaved, has a regular positive solution.
Let us look at what happens at the maximum of $\tilde{\phi}$, which we shall assume occurs at a point $r = r_{max}$. The first two terms in Eq.(\ref{LY3}) will be negative at $r = r_{max}$ so we get
\begin{equation}
\left[\frac{1}{8}A^2\tilde{\phi}^{-7} - \frac{3}{4}\tilde{\phi}^5\right]_{r_{max}} > 0\label{rmax1}.
\end{equation}
This becomes
\begin{equation}
[\max\tilde{\phi}]^{12} < \frac{1}{6}A^2|_{r_{max}} \le \frac{1}{6}\max A^2. \label{maxA}
\end{equation}
The second inequality is needed because the location of $r_{max}$ may well depend on $\alpha$ and, in general, does not coincide with the maximum of $A^2$.  However, Eq.(\ref{maxA}) gives us a uniform upper bound on $\tilde{\phi}$ as a constant independent of $\alpha$.

We get a uniform lower bound by looking at Eq.(\ref{LY3}) when $\tilde{\phi}$ is a minimum, which we  shall assume occurs at $r = r_{min}$. At $r = r_{min}$ we get
\begin{equation}
\frac{1}{6}\min A^2 \le \frac{1}{6} A^2|_{r_{min}} \le 4\frac{[\min\tilde{\phi}]^{8}}{\alpha^4} + [\min\tilde{\phi}]^{12}. \label{minphi}
\end{equation}
Consider the cubic equation
\begin{equation}
x^3 + 4\frac{x^2}{\alpha^4} - \frac{1}{6}\min A^2 = 0.\label{x}
\end{equation}
This will have a positive root that increases as $\alpha$ increases. Fix $\alpha$, $\alpha = \alpha_0$, and find the positive root of Eq.(\ref{x}) for $\alpha = \alpha_0$. This number is a lower bound for $[\min\tilde{\phi}]^4$ that is independent of $\alpha$ for all $\alpha > \alpha_0$. In turn, this means that $\min\phi$ diverges at least as fast as $\alpha$ as $\alpha$ becomes large. Using the bounds on both $\min\phi$ and $\max\phi$, we have shown that there exists a universal constant $C_0$ independent of $\alpha$ such that
\begin{equation}
\frac{\min\phi}{\max\phi} > C_0 \gtrsim \left[\frac{\min A^2}{\max A^2}\right]^{1/3}.
\end{equation}
The maximum and minimum of $\phi$ both increase together proportional to $\alpha $ so that their ratio remains bounded independent of $\alpha$. This can be regarded as a version of the Harnack inequality \cite{GT} for the non-linear equation
Eq.(\ref{LY3}).

\subsection{ $Y < 0,\ \   R < 0$}
We assume that we are in the negative Yamabe class, and set the scalar curvature $R = -24$. Now Eq.(\ref{LY2}) reduces to
\begin{equation}
\nabla^2\tilde\phi + 3\tilde\phi + \alpha^{4}(\frac{1}{8}  A^2 \tilde\phi^{-7} - \frac{3}{4}\tilde\phi^5) = 0.\label{LY4}
\end{equation}
Let us look again at what happens at the maximum of $\tilde{\phi}$. We again assume that this occurs at a point $r = r_{max}$. The first term in Eq.(\ref{LY4}), $\nabla^2\tilde\phi$, will be negative at $r = r_{max}$ so we get
\begin{equation}
\left[\frac{1}{8}A^2\tilde{\phi}^{-7} - \frac{3}{4}\tilde{\phi}^5 + \frac{3\tilde\phi}{\alpha^4}\right]_{r_{max}} > 0\label{rmax2}.
\end{equation}
This becomes
\begin{equation}
[\max\tilde{\phi}]^{12} - \frac{4[\max\tilde{\phi}]^8}{\alpha^4}< \frac{1}{6}A^2|_{r_{max}} < \frac{1}{6}\max A^2. \label{maxA1}
\end{equation}
Consider the cubic equation
\begin{equation}
x^3 - \frac{4x^2}{\alpha^4} - \frac{1}{6}\max A^2 = 0.\label{x2}
\end{equation}
This equation has a positive root which is an upper bound for $[\max\tilde{\phi}]^{4}$ for any given $\alpha$. This root decreases with increasing $\alpha$. An easy way to confirm this is to differentiate Eq.(\ref{x2}) with respect to $\alpha$. We get
\begin{equation}
(3x^2 - \frac{8x}{\alpha^4})\frac{dx}{d\alpha} + \frac{16x^2}{\alpha^5} = 0.
\end{equation}
Using Eq.(\ref{maxA1}), we get $3x^2 - 8x/\alpha^4 = 4x/\alpha^4+ \max A^2/6x > 0.$ This means that $dx/d\alpha < 0$, as required. As in the positive Yamabe case, pick a value for $\alpha$,  $\alpha = \alpha_0$, find the root of Eq.(\ref{x2}), and it will be a uniform upper bound of $\max\tilde{\phi}^4$ for all $\alpha > \alpha_0$.

We get a uniform lower bound by looking at Eq.(\ref{LY4}) at the point where $\tilde{\phi}$ is a minimum. We assume that this occurs at $r = r_{min}$. The first two terms in Eq.(\ref{LY4}) are positive at the minimum. Therefore we get
\begin{equation}
\frac{1}{6}\min A^2 \le \frac{1}{6} A^2|_{r_{min}} \le  [\min\tilde{\phi}]^{12}. \label{minphi1}
\end{equation}
 This inequality Eq.(\ref{minphi1}) gives the desired lower bound for $[\min\tilde{\phi}]$, which is independent of $\alpha$. In turn, this means that $\min\phi$ diverges at least as fast as $\alpha$, as $\alpha$ becomes large. Using the bounds on both $\min\phi$ and $\max\phi$, we have shown that there again exists a universal constant $C_0$ independent of $\alpha$ such that
\begin{equation}
\frac{\min\phi}{\max\phi} > C_0 \gtrsim \left[\frac{\min A^2}{\max A^2}\right]^{1/3},
\end{equation}
and we again recover a Harnack inequality, but now  for
Eq.(\ref{LY4}).

\subsection{$Y = 0,\ \ R = 0$}
Now Eq.(\ref{LY2}) can be reduced to
\begin{equation}
\nabla^2\tilde\phi  + \alpha^{4}(\frac{1}{8}  A^2 \tilde\phi^{-7} - \frac{3}{4}\tilde\phi^5) = 0.\label{LY5}
\end{equation}
Showing the existence of uniform bounds in this case is even easier. We immediately get
\begin{equation}
[\max\tilde{\phi}]^{12} < \frac{1}{6}A^2|_{r_{max}} < \frac{1}{6}\max A^2. \label{maxA2}
\end{equation}
and
\begin{equation}
[\min\tilde{\phi}]^{12} > \frac{1}{6}A^2|_{r_{min}} > \frac{1}{6}\min A^2. \label{minA2}
\end{equation}
Again we get
\begin{equation}
\frac{\min\phi}{\max\phi} > C_0 >  \left[\frac{\min A^2}{\max A^2}\right]^{1/3},
\end{equation}

\subsection{Discussion}
The Harnack inequalities that have been derived in the last three subsections are clearly only valid when $\min(A^2) = \min(K^{ij}_{TT}K^{TT}_{ij}) \ne 0$.  Since $TT$ tensors are usually constructed by a decomposition method, one might think that it would be difficult to find $TT$ tensors which vanish either at points or in regions. Interestingly, we can construct such $TT$ tensors \cite{BaOM} and they cannot be ignored. It has not yet been possible to derive general results for such special TT tensors. Therefore we revert to considering only a spherically symmetric toy model. It turns out, in the spherical situation, that the case where $A^2$ vanishes in a region is easier to analyse than the case where $A^2$ vanishes at a point. When $A^2$ vanishes in a region, we can prove that the minimum of $\phi$ has an upper bound and does not scale with $\alpha$. In the region where $A^2 \ne 0$ we get the standard linear scaling with $\alpha$. We use a combination of analytical and numerical techniques in the next two sections deal with this spherical model.

\section{A Spherical Toy Model: Analytical results}

We will restrict our attention to the case where the base 3-metric is spherically symmetric, and we replace the position-dependent function $A^2$ by an arbitrary spherically symmetric function because we know there exists no regular spherically symmetric $TT$ tensor on flat space or on a spherical compact manifold without boundary\cite{BOM1}. For convenience, we set $K = 3$. We start with a round 3-sphere of constant scalar curvature $R_0$ (a natural choice is $R_0 = 24$) and seek solutions to
\begin{eqnarray}
\nabla^2\phi - \frac{1}{8}R_0\phi + \frac{1}{8}\alpha^{12}A^2\phi^{-7} - \frac{2K^2}{24}\phi^5 &=& 0\\
\nabla^2\phi - 3\phi + \frac{1}{8}\alpha^{12}A^2\phi^{-7} - \frac{3}{4}\phi^5 &=& 0\label{1}
\end{eqnarray}
for varying $\alpha$. It is clear that when either $\alpha \rightarrow 0$ or $\alpha \rightarrow \infty$ strange things happen. In keeping with the focus of this article, we will only consider here the case where $\alpha \rightarrow \infty$.

Looking at Eq.(\ref{1}), it is clear that $\phi$ cannot remain regular as $\alpha \rightarrow \infty$, because the third term would diverge while the others remain regular.  The results derived in Section III remain valid so we know that $\phi$ is linearly proportional to $\alpha$ as long as $A^2$ is nowhere zero.
We need to deal with the situation where $A^2$ vanishes somewhere. In particular,
let us consider the situation where $A^2$ vanishes in a spherical region around the north pole.

 When dealing with spherical symmetry, we are free to take advantage of the fact that a round 3-sphere can be decompactified to flat 3-space, and that conformal transformations form a group under composition. In this picture the equation we wish to study is
\begin{equation}
\nabla^2\hat\phi   + \frac{1}{8}\alpha^{12}\hat A^2\hat\phi^{-7} - \frac{3}{4}\hat\phi^5 = 0,\label{2}
 \end{equation}
with $\hat\phi \rightarrow 0$ at infinity.

The conformal factor that maps flat space into a round sphere of scalar curvature equalling 24 is $\theta = \sqrt{b}/\sqrt{b^2 + r^2}$ for any $b > 0$. The mapping that brings us from Eq.(\ref{1}) to Eq.(\ref{2}) requires that $A^2$ and $\phi$ be transformed. The transformation is that $\hat A^2 = \theta^{-12}A^2$ and $\hat\phi = \theta^{-1}\phi$. In other words, $\phi$ will be finite at the `point at infinity' in the compact manifold ,while $\hat\phi \approx \sqrt{b}/r$ at the corresponding infinity in $R^3$.

  A spherical region around the north pole corresponds to a disc $0 \le r < r_1$, where $r$ is the standard conformally flat coordinate radius, on which $A^2 = 0$. It is now convenient to switch to the asymptotically flat picture. The equation we consider is Eq.(\ref{2}), and, in the disc $0\le r < r_1$ it reduces to
\begin{equation}
\nabla^2\hat\phi    - \frac{3}{4}\hat\phi^5 = 0.\label{2d}
 \end{equation}
 We can write down the solution of this equation explicitly. It is
\begin{equation}
\hat\phi = \frac{\sqrt{a}}{\sqrt{a^2 - r^2}},\label{2e}
\end{equation}
where $a$ is a parameter. These are the functions that map one from flat space to the round hyperboloid of constant negative scalar curvature. These functions blow up at $r = a$, and since we know that the total solution is regular, we know that the blow-up must occur outside the range of validity of these functions. While we do not know {\it a priori} the value of $a$, and it will change with $\alpha$, we do know that we have a lower bound for $a$, i.e., $a \ge r_1$. The minimum value of $\hat\phi$ for these special solutions occurs at the origin. There we get
\begin{equation}
\min\hat\phi = \hat\phi(r = 0) = \sqrt{\frac{1}{a}} \le \sqrt{\frac{1}{r_1}}.
\end{equation}
Therefore $\min\phi$ does not blow up like $\alpha$ but reaches some limit, while $\max\phi$ becomes unboundedly large. A similar argument holds when $A^2$ vanishes near the south pole.

We can repeat this argument when $A^2$ vanishes on some belt $r_1 < r < r_2$. The equation, Eq.(\ref{2d}), is still the same, but now is valid for $r_1 < r < r_2$. We cannot write down a set of explicit solutions, but we have much information about the solutions. Fix the location of the minimum, in this case in the interval $(r_1, r_2)$, and fix the value of $\hat\phi$ at the minimum. This uniquely determines the solution. The solution is `U' shaped, blowing up twice at $r_A$ and $r_B$. The bigger the $\min\hat\phi$, the narrower the `U', i.e., $(r_B - r_A)(\min\hat\phi)^2$ is bounded \cite{CHOM}. However, we know that $(r_A < r_1)$ and $(r_B > r_2)$ because the blow-up cannot occur in the range of validity of Eq(\ref{2d}). Therefore $r_B - r_A > r_2 - r_1$, and so the value of $\min\hat\phi$ is bounded above.

The more interesting case is when $A^2$ vanishes at a point rather than in a region.  To repeat: we can show that when $A^2$ vanishes nowhere, we have uniform blow-up over the entire domain; when $A^2$ vanishes in a region, the minimum saturates. We conjecture that we will have behaviour which is `halfway' between the two situations dealt with above. More precisely, we conjecture that the minimum will blow up with $\alpha$, but at a rate which is slower than linear. The details will depend on the rate that $A^2$ goes to zero at the point. We have done some numerical modeling, both to confirm the analytic results and to investigate those situations where we can prove nothing concrete. These models will be discussed in the next section.

\section{A spherical toy model: numerical results}
In this section, the equation we deal with is a one-dimensional elliptic equation with a highly nonlinear source term,
\begin{equation}
\nabla^2\phi + S(\phi) =0.
\end{equation}
We use a one-dimensional pseudo spectral method to solve this equation in two computational domains \cite{Ansorg,Ansorg2,Pfeiffer}.
We have a coordinate $y \in [0, 2)$ replacing $r \in [0, \infty)$. In the interior domain
we have $y = r, y \in [0, 1],  r \in [0, 1]$, while in the exterior domain we have $y = 2 - 1/r, y \in (1, 2),
r \in (1, \infty)$.
 Therefore, infinity can be included when we put the computational grid at the point $y=2$.

We use Chebyshev polynomials as basis functions and the collocation points are
\begin{eqnarray}
y_{i} &=& \cos [ \frac{\pi  i}{2n} ]\ \  {\rm in}\ \   [0, 1]\\
y_{j} &=& 1 + \cos [ \frac{\pi  j}{2n} ]\ \ {\rm in}\ \ \  [1, 2]
\end{eqnarray}
where $i=0, 1 ...n$ and $j=0,1, 2...n$. The two domains meet at
$i=0$ on the interior domain and $j=n$ on the exterior domain. The
grid point $j=0$ touches the infinity of coordinate $r$.

We require the solution to be $C^1$. This means that $\phi$ and its
normal derivative must match at the interface between the domains.
The discrete equations are solved by the Newton-Raphson method. The
resolution is taken as 100 on each domain.

\subsection{$A^2 = 1$}
We here consider the simplest case, where we choose the base metric to be a round sphere with scalar curvature equalling 24, and pick $A^2$ to be a global constant equalling 1. Then Eq.(\ref{1})
$$
\nabla^2\phi - 3\phi + \frac{1}{8}\alpha^{12}A^2\phi^{-7} - \frac{3}{4}\phi^5 = 0,
$$
 reduces to an algebraic equation
 \begin{equation}
 \phi^{12} + 4\phi^8 - \frac{1}{6}\alpha^{12} = 0,
 \end{equation}
 or in terms of the normalised $\tilde{\phi} = \phi/\alpha$,
 \begin{equation}
 \tilde{\phi}^{12} + \frac{8\tilde{\phi}^8}{\alpha^4} - \frac{1}{6} = 0.
 \end{equation}
 As $\alpha \rightarrow \infty$, we expect $\tilde{\phi}^{12} \rightarrow \frac{1}{6}$ from below, or $\tilde{\phi} \rightarrow 0.8612992$.

 The round base metric can be written as
 \begin{equation}
 g_{ij} = \frac{1}{(1 + r^2)^2}\delta_{ij}.
 \end{equation}
 The computation is all done in the flat space, so we have to solve
 \begin{equation}
\nabla^2\hat\phi   + \frac{1}{8}\alpha^{12}\hat A^2\hat\phi^{-7} - \frac{3}{4}\hat\phi^5 = 0,\label{2c}
 \end{equation}
 where $\hat A^2 = \frac{1}{(1 + r^2)^6}$. At the risk of confusion we introduce a normalized $\hat\varphi = \hat\phi/\alpha$, then Eq.(\ref{2c}) becomes
 \begin{equation}
\nabla^2\hat\phi'   + \frac{1}{8}\alpha^{4}\hat A^2\hat\varphi^{-7} - \frac{3}{4}\alpha^4\hat\varphi^5 = 0.\label{2b}
 \end{equation}
The relationship between $\tilde{\phi}$ and $\hat\varphi$ is  $\tilde{\phi}= \sqrt{1 + r^2}\hat\varphi$.

We solve Eq.(\ref{2b}) numerically with  $\hat A^2 = \frac{1}{(1 + r^2)^6}$. We see that the solution is, as expected, of the form $\hat\phi' = C/\sqrt{1 + r^2}$ with $C$ depending on $\alpha$. In Figure 1 we present $\tilde{\phi}$ on the compact manifold (by multiplying $\hat\phi'$ by $\sqrt{1 + r^2}$). We clearly see that we get a family of constant functions that asymptote to a fixed function as $\alpha$ becomes large.
\begin{figure}[h]
\begin{minipage}[c]{0.5\textwidth}
\centering\includegraphics[width=7cm,height=5cm]{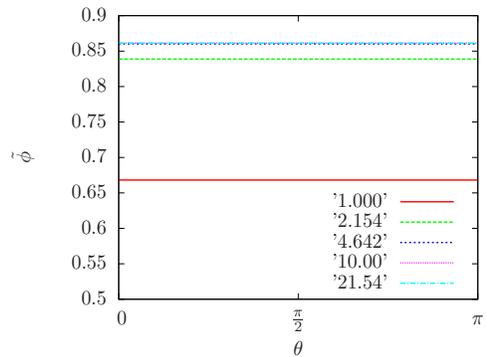}
\end{minipage}
\caption{$\hat\phi$ on $S^3$}
\end{figure}

We list the values of $C$ w.r.t. $\alpha$ as the table I.
\begin{table}
\begin{center}
\begin{tabular}{|c|c|}
\hline
$\alpha$ & C\\
\hline
1 & 0.6680900040565467 \\
\hline
2.154434690 & 0.8387341686204405\\
\hline
4.641588834 & 0.8601789420392659\\
\hline
10 & 0.8612470753749010 \\
\hline
21.54434690 & 0.8612968154854488 \\
\hline
46.41972069 & 0.8612991245657523\\
\hline
\end{tabular}
\end{center}
\caption{It is a list of the scaling parameter $\alpha$ and the
constancy of the standard solution $\hat\phi'$. It is clear that,
as expected, $C$ asymptotes to 0.8612992. }
\end{table}

\subsection{$A^2$  vanishing at a single point}
A simple choice of $\hat A^2$ that corresponds to $A^2$ vanishing at the south pole is to pick $\hat A^2 = \frac{1}{(1 + r^2)^\beta}$ with $\beta > 6$. We restrict our attentions to $\beta=7$ and $\beta=10$.
We present $\hat\phi$ for several values of $\alpha$ with $\beta=7$ and $\beta=10$ in Figure 2 and 3, where $\alpha = 1,2,3,4,5,10$
gives rise to the red, green, blue, pink, light blue, and yellow lines respectively.

\begin{figure}
\begin{center}
\begin{minipage}[c]{0.5\textwidth}
\centering\includegraphics[width=7cm,height=5cm]{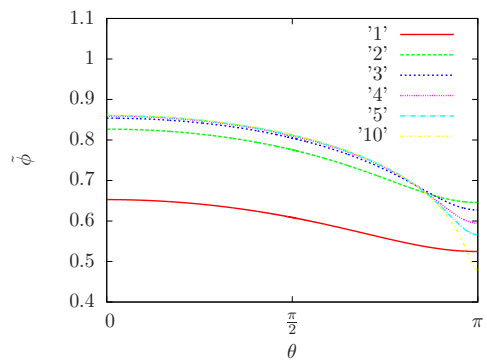}
\caption{$\hat\phi$ on $S^3$ with $\beta=7$ }
\label{}
\end{minipage}
\end{center}
\end{figure}

\begin{figure}
\begin{center}
\begin{minipage}[c]{0.5\textwidth}
\centering\includegraphics[width=7cm,height=5cm]{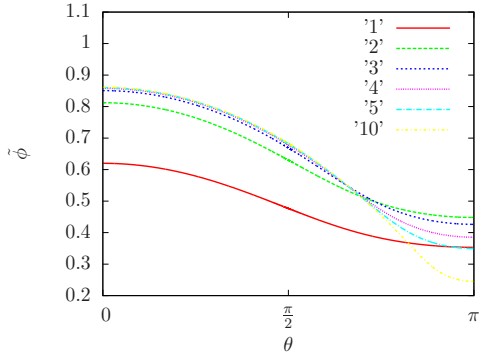}
\caption{ $\hat\phi$ on $S^3$ with $\beta=10$}
\label{}
\end{minipage}
\end{center}
\end{figure}
It is clear that, in both cases, the maximum of $\hat\phi$, at the north pole, settles to a constant value independent of $\alpha$. This shows that the maximum of $\phi$ grows linearly with $\alpha$. On the other hand, the minimum of $\hat\phi$, at the south pole, where $A^2 = 0$, decreases with increasing $\alpha$. Hence the minimum of $\phi$ does not grow linearly with $\alpha$.

To analyse the behaviour of the minimum of $\phi$ as a function of $\alpha$, we consider each of the two cases, i.e.,  $\beta = 7$ and $\beta=10$. We conjecture that $\phi(\theta = \pi)$ scales with some power of $\alpha$, for large $\alpha$. We plot $\ln\phi/\ln\alpha$ versus $\alpha$ for each of the two choices of $\beta$. These are Figures 4 and 5. Each of the two curves flattens out for large $\alpha$. These show that $\min\phi$ grows like $\alpha^n$ with $n \approx 0.71$ when $\beta = 7$, and $\min\phi$ grows like $\alpha^n$ with $n \approx 0.405$ when $\beta = 10$.
\begin{figure}[h]
\begin{center}
\includegraphics[width=7cm,height=5cm]{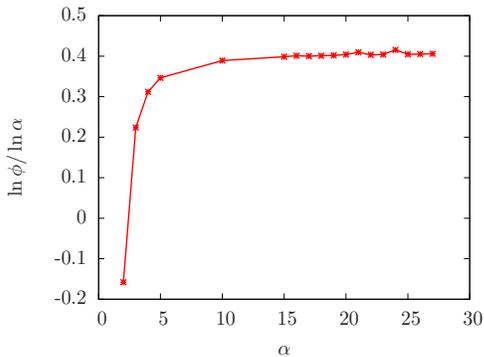}
\end{center}
\caption{$\ln\phi/\ln\alpha$ vs $\alpha$ for $\beta = 7$}
\end{figure}

\begin{figure}[h]
\begin{center}
\includegraphics[width=7cm,height=5cm]{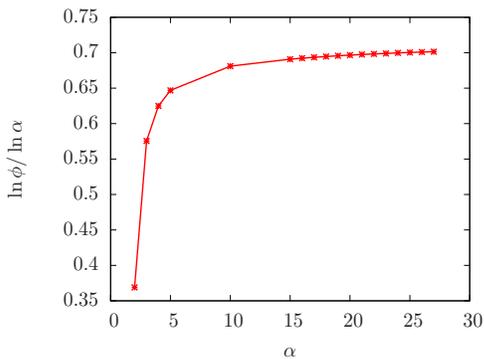}
\end{center}
\caption{$\ln\phi/\ln\alpha$ vs $\alpha$ for $\beta = 10$}
\end{figure}
If $\beta = 6$,  $A^2$ does not vanish at the south pole and we get linear growth of $\phi$ there. The limit $\beta = \infty$ corresponds to $A^2$ vanishing in a region near the south pole, and we get no growth at all. It is nice to see that the growth rate seems to diminish smoothly as we move from $\beta = 6$ to $\beta = \infty$.

\subsection{$A^2$ vanishing in a patch}

We start by finding, on the round sphere, a smooth positive spherical function that vanishes in a region. We will use this as $A^2$.
 We first find a cut function $\eta(x)$ which is defined
 on $[0,2]$ by

 \[ \eta(x) = \left\{ \begin{array}{cc}
 0 & x\in[0,0.5) \\
  209-2240x+10080x^2&\\
  -24640x^3 +35280x^4-29568x^5& \\
  +13440x^6-2560x^7& x\in[0.5,1.0) \\
1&x \in[1.0,\infty)
 \end{array} \right.\]

 This function smoothly interpolates between 0 at $x = 0.5$ and 1 at $x = 1$ and is $C^3$ at each end.
In terms of the rescaled coordinate function $y$ which we introduced at the beginning of this section, we construct the following function
\[ f(y) = \left\{ \begin{array}{cc}
 \eta(1-y) & y\in[0,0.5) \\
 0 & \ \ \ y\in[0.5,1.25) \\
\eta(2y-2)& \ \ \ y\in[1.25,1.5)\\
1& y\in[1.5,2)
 \end{array} \right.\]
 This is a function that is 1 at the origin, goes to zero at $y = r = 0.5$, is zero in $y \in (0.5, 1.25)$, rises to 1 in $y \in (1.25, 1.5)$, and stays equal to unity out to $y = 2, r = \infty$. We choose $\hat{A}^2 = f(y)/(1 + r^2)^6$. This corresponds to a smooth $A^2$ on the sphere, which equals 1 at both the north and south poles but equals zero in a central region. Figure 6 shows $A^2$ on the sphere.

\begin{figure}
\begin{center}
\begin{minipage}[c]{0.5\textwidth}
\centering\includegraphics[width=7cm,height=5cm]{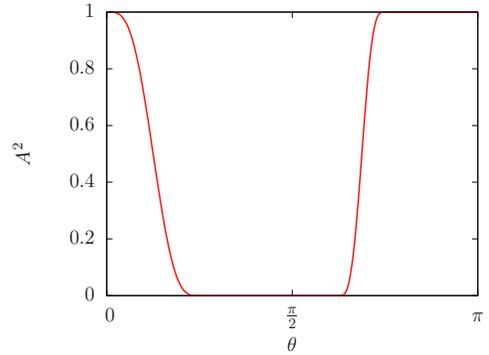}
\caption{$A^2$ on $S^3$}
\label{}
\end{minipage}
\end{center}
\end{figure}

We solve the equation  for a range of parameters $\alpha$. We display the various solutions in Figure 7, showing the normalized $\hat\phi = \phi/\alpha$ on $S^3$.
 The parameters used are $\alpha = 1.000$, $2.154$, $4.642$, $10.00$, $21.54$, $53.13$, $79.37$, and $100.0$. These correspond
 to the red, green, blue, pink, light blue, yellow, dark blue, and mauve lines respectively on Figure 7.

\begin{figure}[h]
\begin{center}
\includegraphics[width=7cm,height=5cm]{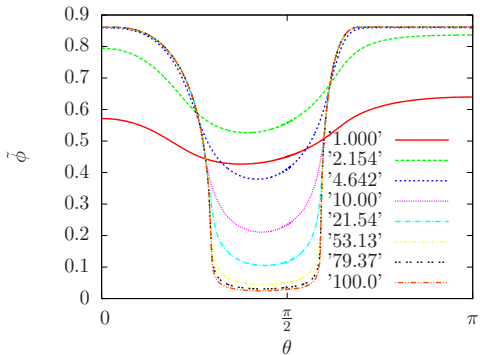}
\end{center}
\caption{$\hat\phi$ on $S^3$ }
\end{figure}
 One
 can clearly see that the value of $\hat\phi$, on the support of $A^2$, tends to a stationary limit, while $\hat\phi$ collapses off the support of $A^2$. This shows that $\phi$ scales linearly with $\alpha$, on the support of $A^2$, while in  the region where $A^2$ is zero, $\hat\phi$ continues to diminish so that $\phi$ approaches a stationary value.

We wish to show that $\min\phi = \alpha\min\hat\phi$ increases with $\alpha$ but approaches some fixed upper bound. The minimum occurs around $\theta = 1.35$. In Figure 8 we have plotted $\ln \phi/\ln \alpha$ versus $\alpha$ at $\theta = 1.35$. One should compare this graph with Figures 4 and 5. This looks like a graph which is going to an asymptotic value of 0, which indicates that $\phi$, at $\theta = 1.35$, is heading for a fixed number, independent of $\alpha$.

\begin{figure}[h]
\begin{center}
\includegraphics[width=7cm,height=5cm]{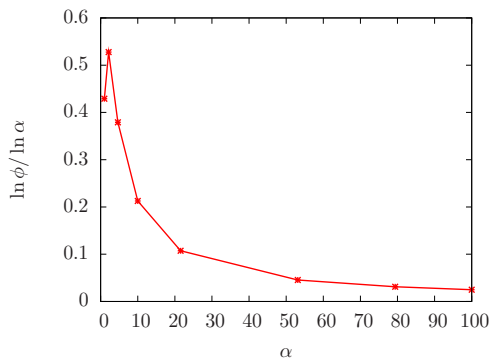}
\end{center}
\caption{$\ln\phi/\ln \alpha$ vs $\alpha$ }
\end{figure}

\section{Conclusions}
We have shown that if we scale $K^{ij}_{TT}K^{TT}{ij}$ by $\alpha^{12}$, we find that the conformal factor, in general, scales like $\alpha$. However, the physical $\bar{K}^{ij}_{TT}\bar{K}^{TT}_{ij} = \phi^{-12}\alpha^{12}K^{ij}_{TT}K^{TT}_{ij} = \tilde\phi^{-12}K^{ij}_{TT}K^{TT}_{ij}$, and as $\alpha$ becomes larger and larger $\tilde\phi$ remains finite. This means that $\bar{K}^{ij}_{TT}\bar{K}^{TT}_{ij}$ remains finite. Hence the velocity part of the gravitational wave energy density remains bounded even though the corresponding `free' data blows up. On the other hand, the volume of space becomes unboundedly large, because $\sqrt{\bar{g}} = \phi^6\sqrt{g} = \alpha^6\tilde\phi^6\sqrt{g}$  blows up. Therefore the total gravitational wave energy in a coordinate sphere becomes larger and larger while the local energy density remains bounded.

We see that the gravitational wave energy inside any coordinate sphere increases like $\alpha^6$ while the surface area increases like $\alpha^4$. This indicates that this family of initial data  will eventually contain horizons.

The numerics, when combined with the analytic calculations, show a coherent picture. If $K^{ij}_{TT}K^{TT}_{ij}$ has no zeros, then the conformal factor blows up uniformly. If the $K^{ij}_{TT}K^{TT}_{ij}$ is zero on a patch, we expect no blow up on this patch, but, nevertheless, we continue to get the standard blowup elsewhere. If $K^{ij}_{TT}K^{TT}_{ij}$ has an isolated zero, we will get blowup at this point, but at a rate slower than in the rest of the space. The rate of blowup is not universal in this case, but depends on how quickly $K^{ij}_{TT}K^{TT}_{ij}$ moves away from zero at that point. There seems to be a smooth transition between the `slow blow-up' case and the `no blow-up' case.

There are a number of obvious extensions to this work. To find a real $TT$ tensor, we need to abandon spherical symmetry and, at the very least, work with axially symmetric data. We would expect that if we have a real $K^{ij}_{TT}K^{TT}_{ij}$, which either vanishes at a point or vanishes in a region,  to get behaviour  similar to the spherical model. We would be surprised if anything could be proven  analytically; we are going to have to depend on numerical modeling. Work is in progress in this direction.

It would be interesting to repeat this analysis in the asymptotically flat case. We would probably want to work with maximal initial data, i.e., $K = 0$, and just have a metric and a $TT$ tensor as free data. It is clear that one can change the metric so that the ADM mass becomes unboundedly large and trapped surfaces appear \cite{BOM}. What happens if we blow-up the extrinsic curvature on a fixed background metric? Will we get the same behaviour? Preliminary investigations indicate that we do: the ADM mass diverges and trapped surfaces appear. We intend to investigate this further.

\begin{acknowledgments}

SB and N\'OM were supported by  Grant 07/RFP/PHYF148 from Science Foundation
Ireland. SB would like to thank Marcus Ansorg and Jos\'e Luis Jaramillo for helpful conversations about the spectral method.

\end{acknowledgments}

\end{document}